# Microalloying effect in ternary Al-Sm-X (X=Ag, Au, Cu) metallic glasses studied by *ab initio* molecular dynamics

J. Xi [1*†], G. Bokas [1,2†], L.E. Schultz [1], M. Gao [1], L. Zhao [1], Y. Shen [1], J.H. Perepezko [1], D. Morgan [1], I. Szlufarska [1,3*]

[1] University of Wisconsin-Madison, Department of Material Science and Engineering, 1509 University Ave., Madison, WI, 53706, U.S.A.
[2] Institute of Condensed Matter and Nanoscience (IMCN), University catholique de Louvain, Louvain-la-Neuve 1348, Belgium
[3] University of Wisconsin-Madison, Department of Engineering Physics, 1500 Engineering Dr., Madison, WI, 53706, U.S.A.

† J. Xi and G. Bokas contributed equally to this work
* Correspondence should be addressed to: szlufarska@wisc.edu, jxi4@wisc.edu

**Abstract**
The icosahedral-like polyhedral fraction (ICO-like fraction) has been studied as a criterion for predicting the glass-forming ability of bulk ternary metallic glasses, $Al_{90}Sm_8X_2$ (X = Al (binary), Cu, Ag, Au), using *ab initio* molecular dynamics (AIMD) simulations. We found that the ICO-like fraction can be determined with adequate precision to explore correlations with AIMD simulations. We then demonstrated that ICO-like fraction correlates with the critical cooling rate, which is a widely used intrinsic measure of glass forming ability. These results suggest that the ICO-like fraction from AIMD simulations may offer a useful guide for searching and screening for good glass formers.

## 1. Introduction

Rapidly quenching metals can lock in a metastable metallic glass (MG) state with many desirable properties, including high elastic modulus, yield strength, specific strength, and good biocompatibility [1]. However, for many applications large MG ingots are needed, where heat removal rates limit cooling rates. It is therefore desirable to have metals with good glass forming ability (GFA) which can be cooled relatively slowly and still form glasses. Of particular interests are MGs that can be cast with more than 1 mm in diameter and which are often called bulk MGs (BMGs). Many metrics exist to quantify GFA, including critical cooling rate ($R_c$), maximum casting diameter ($d_{max}$), the reduced glass transition temperature, $T_{rg} = T_g/T_l$, the supercooled liquid range, $\Delta T_x = T_x - T_g$, and other parameters, like $\gamma = T_x/(T_g + T_l)$ and $\omega = T_g/T_x - 2T_g/(T_g + T_l)$, where $T_g$, $T_x$, and $T_l$ denote the glass transition temperature, the onset of crystallization temperature and the liquidus temperature, respectively. However, the most intrinsic measure of GFA for a material is $R_c$, which is the slowest rate of cooling for which a metal will avoid crystallization and form a glass [1]. Unfortunately, despite many qualitative rules of thumb for guidance and thousands alloy quenching studies, there is still no method to quantitatively predict $R_c$ given just basic compositional information as a starting point.

One promising direction to understand better what controls $R_c$ has been the result that local structure capture by Voronoi Polyhedra (VP), and in particular icosahedral (ICO) polyhedra, can correlate with aspects of GFA [2–5]. VP and related local structural features are important for GFA prediction as they are accessible to molecular simulations, and therefore a correlation with GFA can potentially be coupled to molecular simulations to enable improved prediction of GFA. Of particular interest are two recent studies which directly showed a strong correlation of GFA with ICO properties. Wang *et al.* recently showed that fluctuations in the range of ICO polyhedra seen in molecular dynamics (MD) simulations correlate very well with maximum casting diameter for Al-Ni-Zr alloys [3]. In a similar spirit, Bokas, *et al.* showed that the fraction ICO polyhedra correlates with the ability to form a glass by melt spinning in Al-Sm alloys [4,6]. Both results showed correlation, not a quantitative prediction of some GFA metric, and both focused on just a single or small number of alloys. However, these results raise the possibility that some direct measures of GFA, ideally $R_c$ or $d_{max}$, could be at least to some extent predicted from molecular simulations through the use of VP and fitted correlations. As a further step toward realizing this goal we explore correlations of molecular dynamics determined VP with GFA in a series of previously studied Al-Sm-X ternary alloys.

Our focus on Al-Sm-X alloys is motivated by both the previous success of correlating the GFA with VP from Bokas, *et al.* [4] and the general interest in Al-based MGs. In general, Al-based MG are of significant scientific and engineering interest because of their light weight and corresponding high specific strength, which is larger than most of the aluminum crystalline alloys. Within Al-based rare earth alloys, Al-Sm binary alloys have been reported to have the broadest range of composition with good GFA [7]. Because of that many experimental studies have been performed by melt-quenching Al-Sm [8–12]. Unfortunately, the amorphous matrix of Al-Sm MGs often contains fcc-Al nanocrystals [12,13], suggesting that these alloys have higher $R_c$ than well-known bulk MGs, such as Cu, Mg, and Zr-based. Elevation of the GFA could be accomplished by adding minor alloying elements (known as microalloying) [14–17] to the alloy. For example, ~2 at% substitution of Ni or Y by La or Co in $Al_{86}Ni_8Y_6$ increases the GFA of this alloy, as shown by Yang *et al.* who synthesized experimentally Al-Ni-Y-Co-La MG with the size of approximately 1 mm [18]. Recently, Gao *et al.* published a series of studies of minor alloying in Al-Sm-X alloys exploring their changes in key MG thermo-kinetic properties such as glass transition temperature, $T_g$, and crystallization temperature, $T_x$ [19]. In this work we used this data and explored the ability of VP to correlate with the GFA of these alloys. In particular, we focused on the use of icosahedral fraction (ICF) at room temperature as the relevant VP feature, as this was successfully used in earlier studies [3,4]. Given the previous success of Bokas, *et al.* [4] using VP for predicting GFA in Al-Sm, Al-Sm-X ternaries are a good test bed, as they are close enough to the previous study that similar physics might be expected to apply, but they also significantly expand the domain over which correlations are being explored.

In order to examine trends in GFA with VP for the Al-Sm-X alloys we needed to consider two issues. The first is how to accurately calculate the ICO-like fraction. In particular, to allow for efficient prediction of new alloy properties it is ideal if the ICO-like fraction can be obtained from AIMD, rather than from classical MD with potentials, as the former can be implemented accurately on many more systems since it does not require fitting a potential. Thus, the first part of our work was a careful test of the convergence of ICO-like fraction with respect to simulation times and system size from MD for Al-Sm and Al-Sm-Cu. We note that our successful demonstration that

AIMD can yield adequately converged results is a critical step in enabling the use of local clusters for GFA screening as it readily enables studies across many systems by avoiding the need for potential development for each one. The second issue we considered was the correlation between GFA and ICO-like fraction. To quantify GFA we use $R_c$, which is estimated from a simple cooling model and the thickness of ribbons that yield amorphous systems during melt spinning [20,21] (see Sec. 3.2). We then examined the correlation between $R_c$ and ICO-like fraction for the set of Al-Sm-X glasses. We also considered the correlation between other parameters related to GFA (e.g., $\omega$, $\gamma$, $\Delta T_x$, and $T_{rg}$) and ICO-like fraction.

## 2. Computational methods

To prepare a model of glassy alloys, we first created a face-centered-cubic (fcc) crystal $Al_{92}Sm_8$ system with 256 atoms. The classical MD simulations were performed using Large-scale Atomic/Molecular Massively Parallel Simulator (LAMMPS) code [22]. The glassy alloy was constructed by heating $Al_{92}Sm_8$ crystal to 2000 K and equilibrating it at that temperature via an isothermal-isobaric (NPT) ensemble with the Nose-Hoover thermostat and barostat for 200 ps. The interatomic potential in the formalism of the embedded atom method (EAM) was used [6]. Subsequently, the systems were quenched to 1300 K and equilibrated for another 200 ps under NPT ensemble. This temperature is higher than the liquidus temperature of $Al_{92}Sm_8$. The time step in the classical MD simulations was 1 fs. Atomic structures generated in this process were used as an input for the subsequent AIMD simulations. After that, we substituted Al with X=Ag, Au, and Cu by randomly replacing atoms in the liquid phase to obtain the exact composition, up to 2% of the minor alloying elements. The resulting configurations were further equilibrated in AIMD at 1300 K for 3 ps in the canonical ensemble (NVT) with the Nose-Hoover thermostat and a time step of 3 fs. Afterwards, the systems were cooled down to a set of temperatures with a cooling rate, $5 \times 10^{13}$ K/s. Although the cooling rate can affect the fraction of ICO-like clusters, in the earlier studies, we found that the relative ICO-like fraction among Al-Sm alloys with different compositions does not depend on the cooling rate [4]. Therefore, in the current work, it is reasonable to assume that the cooling rate will not affect the order of the ICO-like fraction in these Al-Sm-X alloys. In addition, we expect that for sufficiently low cooling rates, the ICO-like fraction would saturate near the values that are likely to be found in experimental samples. However, such cooling rates are beyond reach of AIMD simulations because of the computational cost.

The equilibrium volume at particular temperature was obtained by constructing pressure-volume ($P$-$V$) equation of state [23] (see Supplemental Information (SI) Sec. 1). After determining the volume, the samples were equilibrated for at least 3 ps at 300 K, 800 K, and 1300 K with their equilibrium volumes (at 300 K, the samples were equilibrated for 15 ps to get better statistical results). During the processes, 20 runs with different initial atomic configurations were performed for each sample to obtain statistically relevant results. In order to get better statistical results, in these isothermal simulations, the first at least 1.5 ps (at 300 K, the first 12 ps) of the simulation had been treated as the equilibrium run, while the rest of steps was regarded as the production run to calculate the average values of ICO-like fraction. At each temperature, we characterized the atomic structures by counting Voronoi polyhedra. Each cluster can be characterized by the Voronoi indices ($n_3$, $n_4$, $n_5$, $n_6$), where $n_i$ represents the number of cluster faces with $i$ edges. In particular, the perfect ICO clusters contain only pentagon polyhedra in their structure, and thus their Voronoi indices are (0, 0, 12, 0). Voronoi polyhedra that include 8 or more faces with 5 edges each and have a coordination number between 11 and 13 inclusive are defined as ICO-like clusters [24–26]. Calculation was done using the OVITO software with an edge threshold of 0.1 [4]. The edge threshold of a VP cluster is defined as the minimum length an edge of a face can have.

AIMD simulations were performed in the framework of the density functional theory using Vienna Ab-Initio Simulation Package (VASP) [27]. Projector-augmented-wave (PAW) potentials [28] were used to mimic the ionic cores, while the generalized gradient approximation (GGA) in the Perdew-Burke-Ernzerhof (PBE) [29] approach was employed for the exchange and correlation functional. All computations were performed with a cutoff energy of 1.2 times of the maximum cutoff (ENMAX) for the plane wave basis set (i.e., 288.48 eV, 299.81 eV, 288.48 eV, and 327.84 eV for $Al_{92}Sm_8$, $Al_{90}Sm_8Ag_2$, $Al_{90}Sm_8Au_2$, and $Al_{90}Sm_8Cu_2$, respectively). The larger energy cutoff, 400 eV, has been tested for $Al_{92}Sm_8$ and $Al_{90}Sm_8Au_2$, and the values of ICO-like fraction in both systems are comparable to these in the lower energy cutoff cases, suggesting that the 1.2 times of ENMAX is enough to simulate the icosahedral-like polyhedral. The integration over the Brillouin zone was performed using the $\Gamma$ point for AIMD simulations.

Convergence tests were performed in $Al_{92}Sm_8$ and $Al_{90}Sm_8Cu_2$ by using the classical MD simulations to determine expected errors in the averages calculated using the system size and simulation times accessible to AIMD simulations. In the classical MD simulations, three different system sizes (containing 256, 2048, and 6912 atoms, corresponding to approximately cubic supercells of 1×1×1, 2×2×2, and 3×3×3) were tested. These glassy alloys were first equilibrated using a constant pressure and constant temperature (NPT) ensemble for 100 ps at a given temperature (300 K, 800 K, and 1300 K) while maintaining zero pressure. After that, the equilibrium volume at each temperature was determined so that this volume could be used in subsequent runs in the NVT ensemble. To determine the ICO-like fractions at different temperatures, 100 independent NVT runs were carried out, where independence was assured by starting the runs from different initial velocity distributions. For each NVT run, isothermal holds were done at 300 K, 800 K, and 1300 K for 15 ps each. For each frame in each independent run, the ICO-like fractions were calculated by counting the ICO-like VP for each temperature hold and then dividing by the total number of VP present in the system. The first 5 ps of the isothermal simulation had been treated as the equilibrium run, while the next 10 ps was the production run, at which we calculated the average values of ICO-like fraction. The time step in the classical MD simulations was 1 fs.

## 3. Results and Discussion

### 3.1 Convergence testing

To enable practical AIMD simulations it is advantageous to use simulation cells of 256 atoms and perform at most about 20 repeated simulation runs. To evaluate possible size effects and confirm that the simulation cell with 256 atoms is large enough, the fractions of ICO-like clusters in $Al_{92}Sm_8$ and $Al_{90}Sm_8Cu_2$ systems were calculated via classical MD simulations using three different simulation cell sizes (256, 2048, and 6912 atoms). The values of ICO-like fraction in these systems were plotted as a difference from the ICO-like fraction in the system with 6912 atoms. The raw data can be found in Table S2. Specifically, if the ICO-like fraction in a system with $N$ atoms is $f_N$, then the quantity we plot can be written as $\Delta f = (f_N - f_{6921})$. As shown in Fig. 1 for $Al_{92}Sm_8$ and $Al_{90}Sm_8Cu_2$, $\Delta f$ is smaller than 1.0%. In addition, previous AIMD simulations have shown that a relatively small size system with 256 atoms can provide us reasonable values of kinetic properties of Al-Sm alloys, such as the diffusion coefficient [4]. Therefore, in order to balance the computational cost and the accuracy of the results, we selected the 256 atoms for our AIMD simulations.

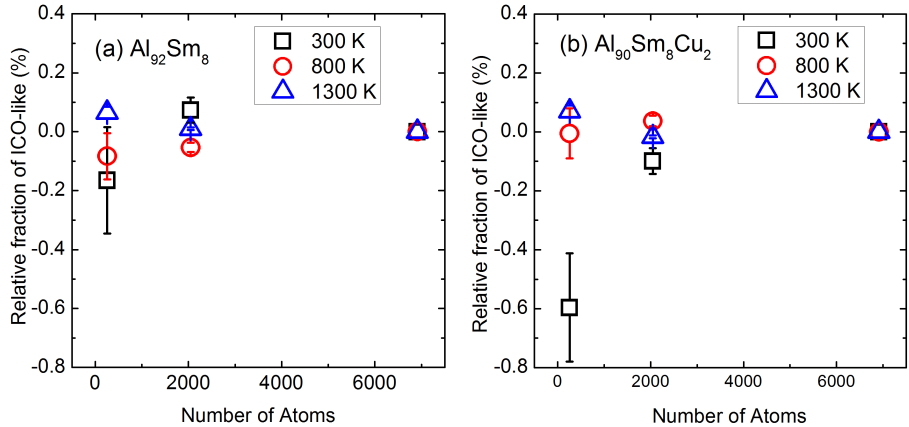

*Fig. 1. The ICO-like fraction difference ($\Delta f$) as a function of the system size in (a) $Al_{92}Sm_8$ and (b) $Al_{90}Sm_8Cu_2$. $\Delta f$ is referenced to the ICO-like fraction, $f_{6921}$ in the system with 6921 atoms. The raw values for ICO-like fractions are listed in Table S2.*

To demonstrate that about 20 AIMD simulation runs were adequate for converged results, we calculated how the standard error in the mean (SEM) of ICO-like fraction depends on the number of runs. Here, we mainly focused on the low temperature runs at 300 K for the smallest supercell of $Al_{92}Sm_8$ and $Al_{90}Sm_8Cu_2$ with 256 atoms, as shown in Fig. 2. SEM is determined as the standard deviation of the ICO-like fractions divided by the square root of the number of runs included in an average. For each point in Fig. 2, values from individual runs can be added to the average in different order, which results in slightly different convergence of SEM with the number of runs. The error bars shown in Fig. 2 are the standard deviation of the mean SEM values obtained using 100 random shuffling orders of values determined from individual runs. Hence, Fig. 2 shows how uncertainty decreases as a function of the number of runs and the kind of spread that is expected in that estimated uncertainty. We include results from both the 100 runs with interatomic potentials and 20 runs with AIMD. From Fig. 2, we can see that the SEM of ICO-like fraction becomes smaller than about 0.6% when the number of simulation runs is larger than 20, which suggested that the number of AIMD runs with different initial atomic configurations are statistically significant.

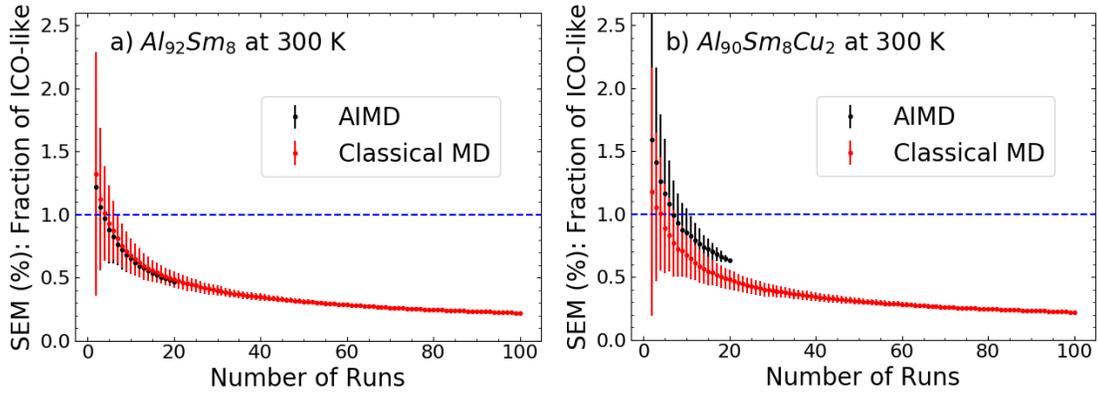

*Fig. 2. Convergence of the standard error in the mean (SEM) for the fraction of ICO-like clusters in $Al_{92}Sm_8$ and $Al_{90}Sm_8Cu_2$ as a function of the number of MD runs at 300 K. The dashed line corresponds to a SEM equal to 1%.*

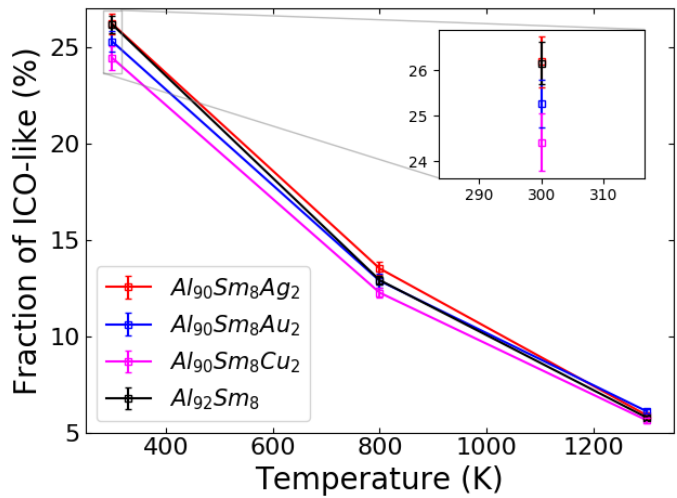

*Fig. 3. The ICO-like fraction of the listed alloys as a function of temperature.*

The effect of temperature on the magnitude and SEM of ICO-like fraction of $Al_{92}Sm_8$, $Al_{90}Sm_8Ag_2$, $Al_{90}Sm_8Au_2$, and $Al_{90}Sm_8Cu_2$ was also considered, as shown in Fig. 3 and Fig. 4, respectively. It was found that the magnitude and SEM of ICO-like fraction of these alloys are decreased as temperature increases. These reductions with temperature are understandable as the greater thermal disorder spreads the system out over more VP types, lowering the magnitude of the ICO-like fraction. This lowering of magnitude, combined with reduced molecular dynamics correlation times (i.e., more averaging) at higher temperatures, leads to a lower SEM of ICO-like fraction. In fact, the ICO-fraction SEM decreases faster than the magnitude, leading to a reduction in the percent uncertainty of the ICO-fraction with increasing temperature (see Fig. 4).

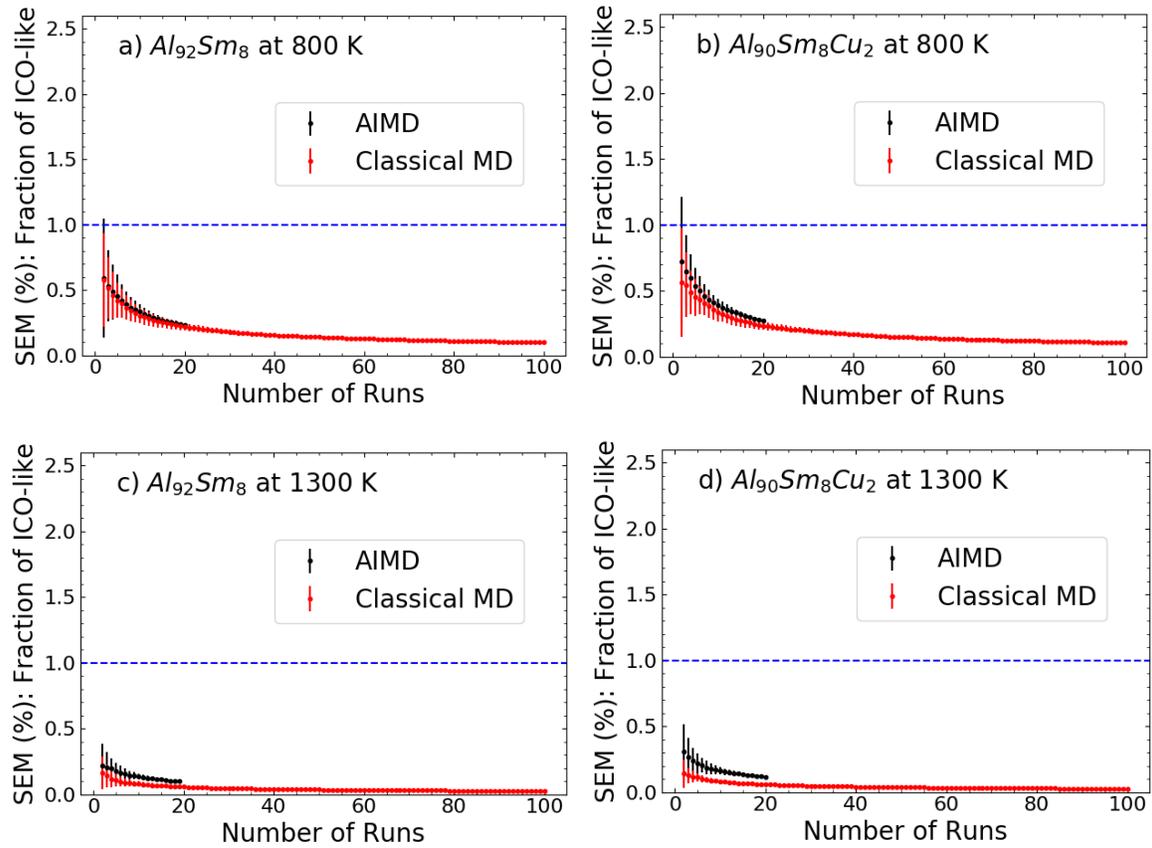

*Fig. 4. Convergence of the standard error in the mean (SEM) for the fraction of ICO-like clusters in $Al_{92}Sm_8$ and $Al_{90}Sm_8Cu_2$ as a function of the number of MD runs at 800 K and 1300 K. The dashed line corresponds to SEM equal to 1%.*

### 3.2 Correlation of ICO-like fraction with GFA criteria

*Table 1. Experimentally measured characteristic thermal parameters (in Ref. [19][31]): glass transition temperature $T_g$, primary crystallization temperature $T_x$, liquidus temperature $T_l$, the range for critical thickness, $L_c$, and the corresponding critical cooling rate, $R_c$.*

| Sample | $T_g$ (K) | $T_x$ (K) | $T_l$ (K) | $L_c$ (μm) | $R_c$ ($10^5$ K/s) |
|---|---|---|---|---|---|
| $Al_{92}Sm_8$ | 363 | 441 | 1138 | 74-110 | 1.59-3.56 |
| $Al_{90}Sm_8Cu_2$ | 357 | 396 | 1108 | 42-54 | 6.40-10.71 |
| $Al_{90}Sm_8Ag_2$ | 366 | 434 | 1143 | 78-130 | 1.14-3.21 |
| $Al_{90}Sm_8Au_2$ | 371 | 462 | 1224 | 50-63 | 5.34-8.58 |

As a next step we examine the possible correlation between the ICO-like fraction and $R_c$. The magnitude of $R_c$ is predicted based on the approximate relationship between cooling rate $R$, sample thickness, and thermodynamic properties, $R = (T_l - T_g)K / CL^n$ [20,21], where $L$ is the thickness of the sample, $K$ and $C$ are the thermal conductivity and the heat capacity per unit volume, respectively. The index $n$ is a constant that is related to the rate of heat transfer between the sample and its surroundings; $n$ takes values between 1 and 2 [30]. If the heat transfer out of the system is limited by the heat flow within the alloy melt (rather than by the heat flow from the surface into the surroundings), then the index $n$ would be close to 2. This condition is expected to hold during our melt spinning experiments [31], and therefore, in this work, we assumed that the index $n$ is equal to 2. Any similar value of $n$ will yield the same order of $R_c$ values and therefore it will not impact the qualitatively conclusions in this paper.

The critical cooling rate, $R_c$, corresponds to the largest possible thickness that can be made and still form a glass, which we call the critical thickness, $L_c$. We take $K\sim 0.1$ J·cm$^{-1}$·s$^{-1}$·K$^{-1}$ (typical of a molten alloys), and $C\sim 4$ J·cm$^{-3}$·K$^{-1}$ (also typical of molten alloys) [21]. While these approximate values will introduce errors in our $R_c$ estimates, these errors will likely be very similar across all the materials due to their similar compositions. Therefore, these errors approximately represent an unknown scale factor in the value of $R_c$ and will not impact any of the qualitative trends shown in this work. The critical cooling rate can then be determined by $L_c$ and the thermodynamic properties $T_l$ and $T_g$. The thickness was measured in previous melt-spinning experiments along with whether the alloy was amorphous or not [31]. The thickness was measured only at certain locations so the exact value of $L_c$ was not measured. Therefore, we give a range for the $L_c$ (see Table 1), where the upper bound of the range corresponds to the smallest $L$ where the sample begins to have mixed amorphous and crystalline phases, and the lower bound of the range corresponds to the largest $L$ where the sample has only an amorphous phase. The true $L_c$ corresponding to the critical cooling rate should be within this range. We estimate a precise value for $L_c$ as the midpoint of the range, which is what is shown for $L_c$ in Fig. 5(a). $T_l$ and $T_g$ were obtained from previous experimental measurements [19] and are listed in Table 1. These values are then used to determine $R_c$ and the correlation between the $R_c$ and the AIMD simulated ICO-like fraction is shown in Fig. 5.

As shown in Fig. 5, there is a variation of ICO-like fraction in systems with different alloying elements, X. This variation could be due to the atomic radius and/or chemical bonding. In order to shed light on the reasons underlying the change in icosahedra fraction, we have analyzed the relationship between the fraction ICO-like clusters and the average coordination number, the bond distance, the atomic radius, and the strength of the chemical bonding between the alloying elements (X) and Al. The chemical bond strength has been analyzed by calculating force constants and vacancy formation energies. The reference state for the calculation of vacancy formation energy is an isolated X atom because we want to identify the effect of bond strength, rather than the true formation energy for element X.

The results are summarized in Fig 6 and in Fig. S2. Specifically, in Fig. 6(a), we can see that there is a monotonic relationship between the ICO-like fraction and the average coordination number of the alloying elements considered in our study. A similar trend can also be observed between the ICO-like fraction and the Al-X bond distance (see Fig. S2(a)). The coordination number and the bond distance can be affected by the atomic size and by the strength of the bonds. It turns out that both of these factors play a role and their effects cannot be separated from each other. In particular, in Fig. 6(b) we plot the relationship between ICO-like fraction and the atomic radius of X, and although the atomic size of Ag and Au are comparable to each other [32], the ICO-like fractions are different. Consequently, the atomic size alone does not explain the trend in the ICO-like fraction in these alloys. To determine the influence of the Al-X bond strength, we have calculated the force constant on the atom X (see Fig. 6(c)) as well the formation energy of X vacancy in the Al-Sm-X systems (see Fig. S2(b)). Again, the relationship is not monotonic, which means that the chemical bond strength alone cannot explain the trend in ICO-like fraction.

We have also found that the alloying element X affects the composition of the ICO-like clusters. These results are not central to the conclusions shown in this paper and therefore analysis of cluster composition is reported in SI Sec. 4.

From Fig. 5(b), we can see that the values of $\log_{10}(R_c)$ for $Al_{92}Sm_8$ and $Al_{90}Sm_8Ag_2$ are close to each other, but obviously smaller than those for $Al_{90}Sm_8Au_2$ and $Al_{90}Sm_8Cu_2$. The similarity between $\log_{10}(R_c)$ for $Al_{90}Sm_8Au_2$ and $Al_{90}Sm_8Cu_2$ makes it difficult to argue that the value of $\log_{10}(R_c)$ for $Al_{90}Sm_8Au_2$ is statistically significantly smaller than that for $Al_{90}Sm_8Cu_2$. The difference may be correct and reproducible, but more work is needed for confirmation. By comparing the ICO-like fraction and $\log_{10}(R_c)$ for these samples, we can see that there is a correlation between critical cooling rate and ICO-like fraction. This result suggests that the correlation between the ICO-like fraction and GFA that has been previously shown to exist in binary Al-Sm alloy [4,6], is likely also present in the corresponding ternary Al-Sm-X alloys, although more data is needed to establish such a correlation in general. In addition, it is worth noting that the above results demonstrate that AIMD simulations can be used successfully to predict metallic glasses with good GFA, based on known structural descriptors. This is a useful finding since *ab initio* MD simulation can explore alloys with different compositions without the need to fit a classical force field to each of the compositions. We have also considered the correlation between other parameters related to GFA (specifically, $\gamma$, $\Delta T_x$, $T_{rg}$) and ICO-like fraction. We consider these less robust measures of GFA and have therefore put the relevant figures and discussion in the SI Sec. 5.

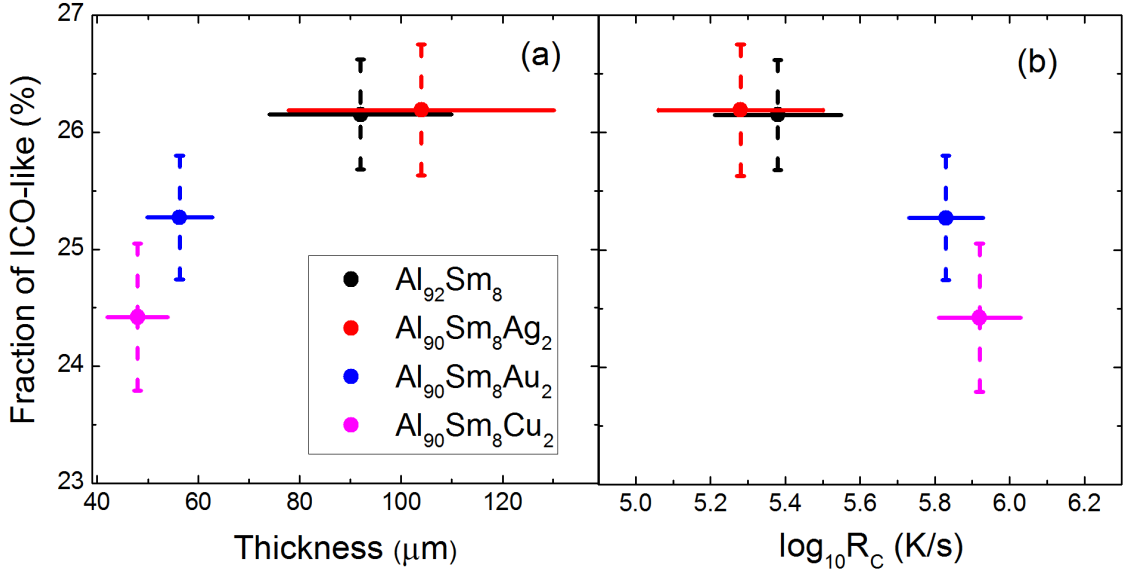

*Fig. 5. The ICO-like fraction for $Al_{92}Sm_8$, $Al_{90}Sm_8Ag_2$, $Al_{90}Sm_8Au_2$, and $Al_{90}Sm_8Cu_2$ as a function of the experimentally measured critical thickness (a) and estimated critical cooling rate $R_c$ (b). The solid line represents the range of experimentally measured thicknesses bounding $L_c$ and the corresponding range for cooling rate. The dashed line represents the error bar from AIMD simulations.*

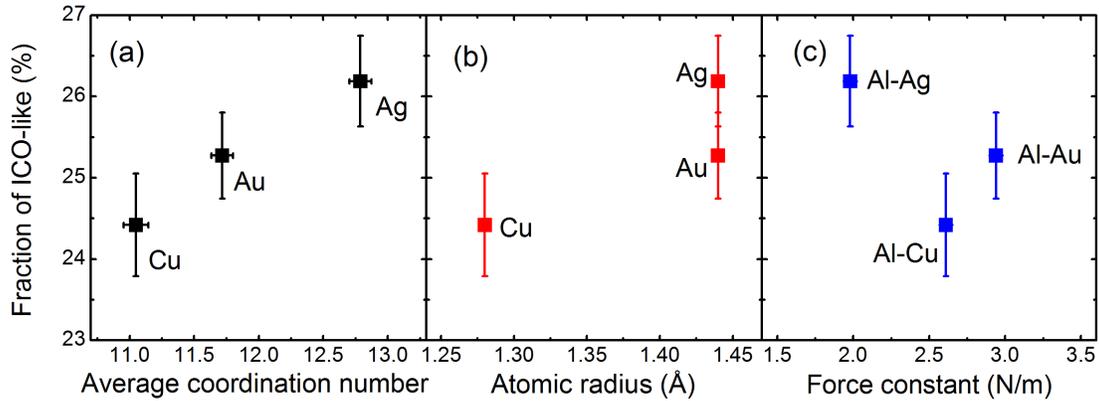

Fig. 6. The ICO-like fraction as a function of (a) the average coordination number of X, (b) the atomic radius of X, and (c) the force constant of Al-X in $Al_{90}Sm_8Ag_2$, $Al_{90}Sm_8Au_2$, and $Al_{90}Sm_8Cu_2$ systems. Here X denotes the alloying element of Cu, Au, and Ag. The error bar in the figure is the standard error of mean.

**Conclusions**

This work showed two results that will aid in the development of simulation accessible features related to VP for predicting GFA. First, we demonstrated that the room temperature ICO-like fraction of a quenched glass can be determined to under 1% precision with system sizes of 256 atoms and averaging over approximately 20 runs of about 3 ps each. These systems sizes, times, and run quantities demonstrate that this feature is accessible to direct AIMD, and therefore can be readily calculated for many systems. Second, we demonstrated that an estimated critical cooling rate, $R_c$, correlated well with ICO-like fraction across a set of four $Al_{90}Sm_8X_2$ (X = Al (binary), Cu, Ag, Au) alloys. This result is consistent with the hypothesis that the true $R_c$, and therefore GFA, correlates with ICO-like fraction in this family of alloys. These studies provide guidance on methods to determine ICO-like fraction from AIMD and support the further exploration of this feature as a guide for in-silico design of MGs with good GFA.

**Acknowledgements**

We thank V. Jambur for useful discussions. We also gratefully acknowledge support from the National Science Foundation (NSF), Division of Materials Research (DMR), under Award #1728933. Computational support was provided by the Extreme Science and Engineering Discovery Environment (XSEDE), which was supported by the National Science Foundation Grant No. OCI-1053575.

# Supplemental materials

## 1. Determination of equilibrium volume

The systems were cooled down from 1300 K to 1100 K, 900 K, 700 K, 500 K, and 300 K with a cooling rate, $5 \times 10^{13}$ K/s, using AIMD simulations. At each particular temperature, the systems with a set of initial-guessed volumes were run for 3 ps with the NVT ensemble. The pressure of each run was obtained, and the equilibrium volume was determined by constructing pressure-volume (*P-V*) equation of state [1] and finding the equilibrium volume associated with zero pressure. The equilibrium volumes at these temperatures are summarized in Table S1. The equilibrium volume at 800 K was linearly interpolated from fitting all the volumes at different temperatures. Here we should note that although the approximated volume at 800 K may be different from the real equilibrium volume at that temperature due to the glass transition within that temperature range, the deviation in equilibrium volume at 800 K should be very small and not affect our conclusions.

*Table S1. The equilibrium volume at the particular temperature for different alloys.*

| Temperature (K) | Equilibrium volume with 256 atoms (Å³) | | | |
|---|---|---|---|---|
| | $Al_{92}Sm_8$ | $Al_{90}Sm_8Ag_2$ | $Al_{90}Sm_8Au_2$ | $Al_{90}Sm_8Cu_2$ |
| 1300 | 5359.38 | 5422.09 | 5360.29 | 5441.56 |
| 1100 | 5257.95 | 5352.78 | 5289.43 | 5273.67 |
| 900 | 5226.46 | 5241.15 | 5226.21 | 5194.96 |
| **800** | **5132.95** | **5149.04** | **5133.85** | **5167.85** |
| 700 | 5124.03 | 5116.37 | 5117.23 | 5086.63 |
| 500 | 4980.49 | 4994.56 | 5010.52 | 4980.56 |
| 300 | 4920.97 | 4905.19 | 4920.87 | 4891.44 |

## 2. Correlation of ICO-like fraction with atomic size and chemistry of alloying element

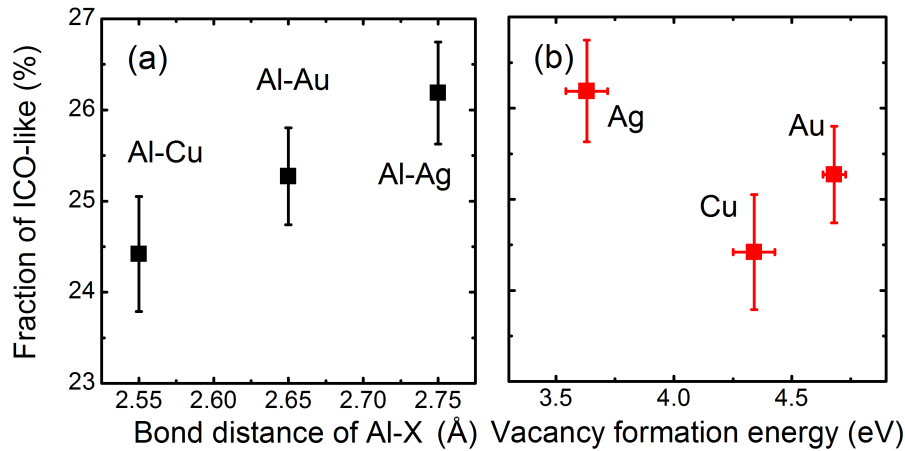

*Fig. S2. The ICO-like fraction as a function of (a) the bond distance of Al-X, (b) the vacancy formation energy of X in $Al_{92}Sm_8$, $Al_{90}Sm_8Ag_2$, $Al_{90}Sm_8Au_2$, and $Al_{90}Sm_8Cu_2$. Here X denotes the alloying element of Cu, Au, and Ag. The error bar in the figure is the standard error of mean.*

## 3. The size effect on ICO-like fraction in $Al_{92}Sm_8$ and $Al_{90}Sm_8Cu_2$ systems

Table S2. The ICO-like fraction, f, in $Al_{92}Sm_8$, $Al_{90}Sm_8Cu_2$ systems with different system sizes and at different temperatures. The value in parenthesis is the standard error of the mean from 100 classical MD runs.

| System size [atom] | $f(Al_{92}Sm_8)$ [%] Temperature [K] | | | $f(Al_{90}Sm_8Cu_2)$ [%] Temperature [K] | | |
|---|---|---|---|---|---|---|
| | 300 | 800 | 1300 | 300 | 800 | 1300 |
| 256 | 28.36(0.22) | 17.69(0.09) | 6.69(0.02) | 26.91(0.22) | 17.32(0.11) | 6.66(0.02) |
| 2048 | 28.60(0.08) | 17.72(0.04) | 6.63(0.01) | 27.40(0.08) | 17.36(0.04) | 6.58(0.01) |
| 6912 | 28.52(0.04) | 17.77(0.02) | 6.62(0.01) | 27.50(0.04) | 17.32(0.02) | 6.59(0.01) |

## 4. Chemical composition of ICO-like clusters with Al and X atoms at the center in $Al_{90}Sm_8Cu_2$, $Al_{90}Sm_8Au_2$, and $Al_{90}Sm_8Ag_2$ systems

Table S3. The average chemical composition and the average coordination number (CN) of ICO-like clusters with Al at the center in $Al_{90}Sm_8Cu_2$, $Al_{90}Sm_8Au_2$, and $Al_{90}Sm_8Ag_2$ systems from AIMD simulations. Here, the value in parenthesis is the standard error of the mean based on 20 AIMD runs.

| | # of Al | # of Sm | # of X | CN |
|---|---|---|---|---|
| $Al_{90}Sm_8Ag_2$ | 11.15 (0.01) | 1.21 (0.01) | 0.19 (0.01) | 12.55 |
| $Al_{90}Sm_8Au_2$ | 11.15 (0.01) | 1.23 (0.01) | 0.16 (0.01) | 12.54 |
| $Al_{90}Sm_8Cu_2$ | 11.12 (0.01) | 1.26 (0.01) | 0.17 (0.01) | 12.54 |
| $Al_{92}Sm_8$ | 11.32 (0.01) | 1.23 (0.01) | / | 12.54 |

Table S4. The average chemical composition and the average coordination number (CN) of ICO-like clusters with X atom at the center (X=Ag, Au and Cu) in $Al_{90}Sm_8Cu_2$, $Al_{90}Sm_8Au_2$, and $Al_{90}Sm_8Ag_2$ systems from AIMD simulations. The value in parenthesis is the standard error of the mean from 20 AIMD runs.

| | # of Al | # of Sm | # of X | CN |
|---|---|---|---|---|
| $Al_{90}Sm_8Ag_2$ | 9.16 (0.14) | 1.87 (0.09) | 1.14 (0.05) | 12.17 |
| $Al_{90}Sm_8Au_2$ | 8.67 (0.17) | 1.81 (0.13) | 1.05 (0.03) | 11.53 |
| $Al_{90}Sm_8Cu_2$ | 8.75 (0.13) | 1.40 (0.09) | 1.13 (0.05) | 11.28 |

We have analyzed the chemical composition of the ICO-like clusters with Al at the center and with an X atom at the center, as shown in Table S3 and Table S4, respectively. Sm-centered clusters were not analyzed because we have not found any ICO-like Sm-centered clusters in the alloys. For the Al-centered ICO-like cluster, it is found that the existence of an X atom can slightly decrease the number of Al atoms within the cluster, while the changes in the number of Sm is negligible. If we look at the X-centered clusters (see Table S4), the chemical composition distribution is quite different from that of the Al-centered clusters. For example, we can see that the number of Al atoms within the X-centered cluster is ~9, which is smaller than that in the Al-centered cluster (~11). However, the numbers of Sm and X atoms within the X-centered cluster are larger than those in the Al-centered cluster. In addition, we found that the $Al_{90}Sm_8Ag_2$ system has the largest number of Al within the ICO-like cluster, which is followed by the $Al_{90}Sm_8Cu_2$ system, and then by the $Al_{90}Sm_8Au_2$ system. This trend is consistent with the order of the bond strength, as shown in Fig. 6(c) in the main text. Furthermore, it is interesting to mention that the order of the average coordination number of X atoms within X-centered clusters is consistent with the bond distance of Al-X, as shown in Fig. 6(a) of the main text and Fig. S2(a). This finding is consistent with the previously reported correlation between the bond distance and the coordination number [5].

## 5. Correlation of ICO-like fraction with other potential GFA criteria

The AIMD simulated ICO-like fractions in these alloys have also been correlated with other potential GFA criteria. These include $\gamma=T_x/(T_g+T_l)$, $\Delta T_x=T_x-T_g$, and $T_{rg}=T_g/T_l$, and $\omega=T_g/T_x-2T_g/(T_g+T_l)$ (where $T_g$, $T_x$, and $T_l$ denote the glass transition temperature, the onset of crystallization temperature and the liquidus temperature, respectively, which were measured from previous experiments [3]). These are shown in Fig. S3, where we also include the correlations with $R_c$ (see main text) for completeness. The correlation with $\gamma$ is similar to that with $\omega$, as might be expected since $\gamma$ and $\omega$ tend to correlate quite strongly [4]. The weak correlation with $T_{rg}$ and $\Delta T_x$ is perhaps not surprising as these variables are only loosely correlated with GFA. The correlation with $R_c$ appears to be the best of the all these GFA metrics, which is encouraging as $R_c$ is the only true direct measure of GFA.

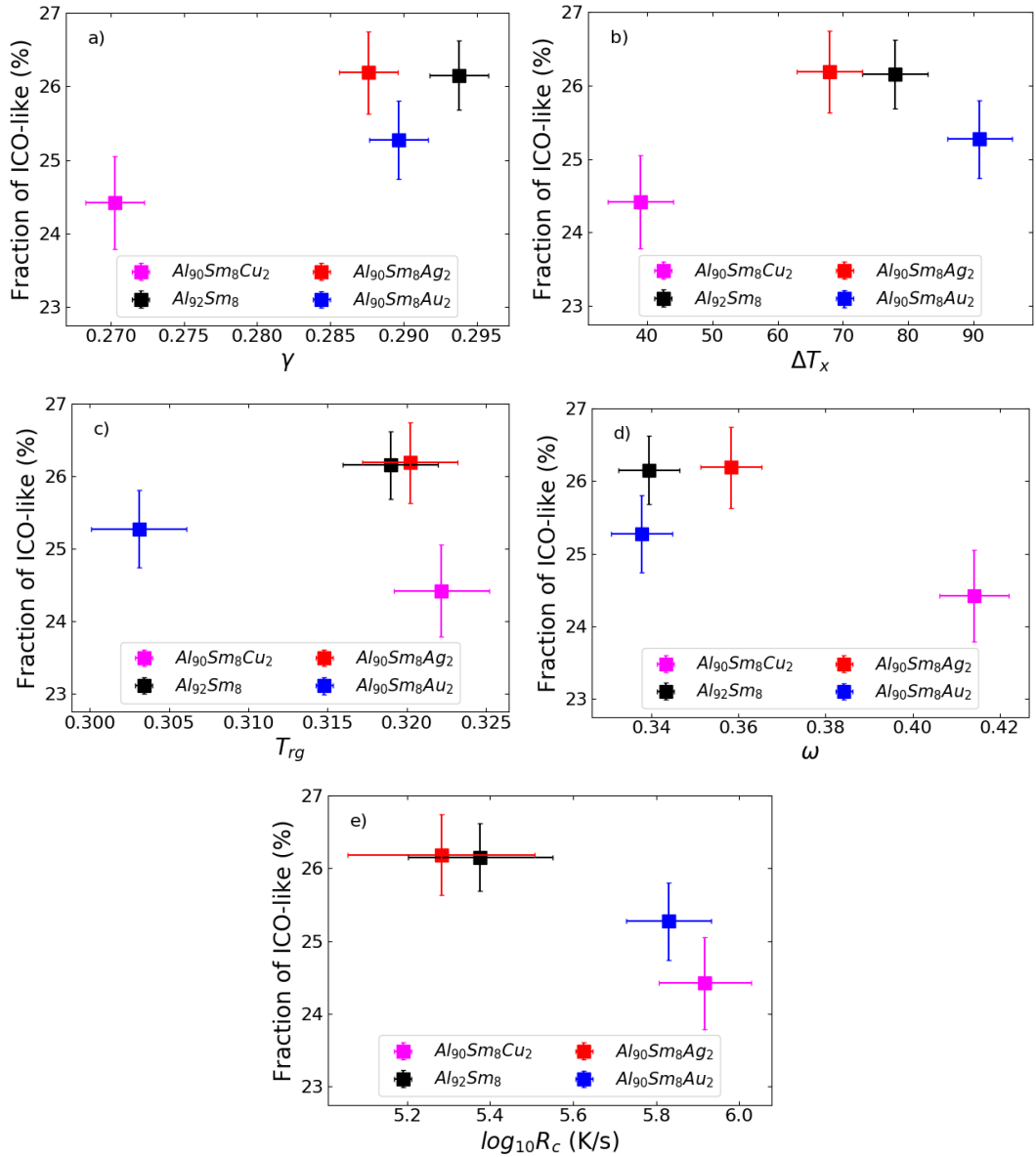

*Fig. S3. The ICO-like fraction as a function of the GFA criteria, like (a) γ, (b) ΔT$_x$, (c) T$_{rg}$, (d) ω, and (e) R$_C$, for Al$_{92}$Sm$_8$, Al$_{90}$Sm$_8$Ag$_2$, Al$_{90}$Sm$_8$Au$_2$, and Al$_{90}$Sm$_8$Cu$_2$.*